\documentstyle[12pt,citesort,epsfig]{dubna98}
\begin{document}
\newcommand{\eq}{\begin{eqnarray}}
\newcommand{\en}{\end{eqnarray}}
\newcommand{\Ms}{M^\star}
\newcommand{\Ps}{P^\star}
\newcommand{\Es}{E^\star}
\newcommand{\mc}{m_\pi^2}
\newcommand{\mo}{m_{\pi^0}^2}
\newcommand{\dpi}{\Delta_\pi}

\begin{figure}
BUHE 98-09 (20 August 1998):\\
Contribution  to the International Workshop "Hadronic Atoms\\  
and Positronium in the Standard Model", Dubna 26-31 May 1998 

\begin{center}
\vspace{1cm}
\epsfig{file=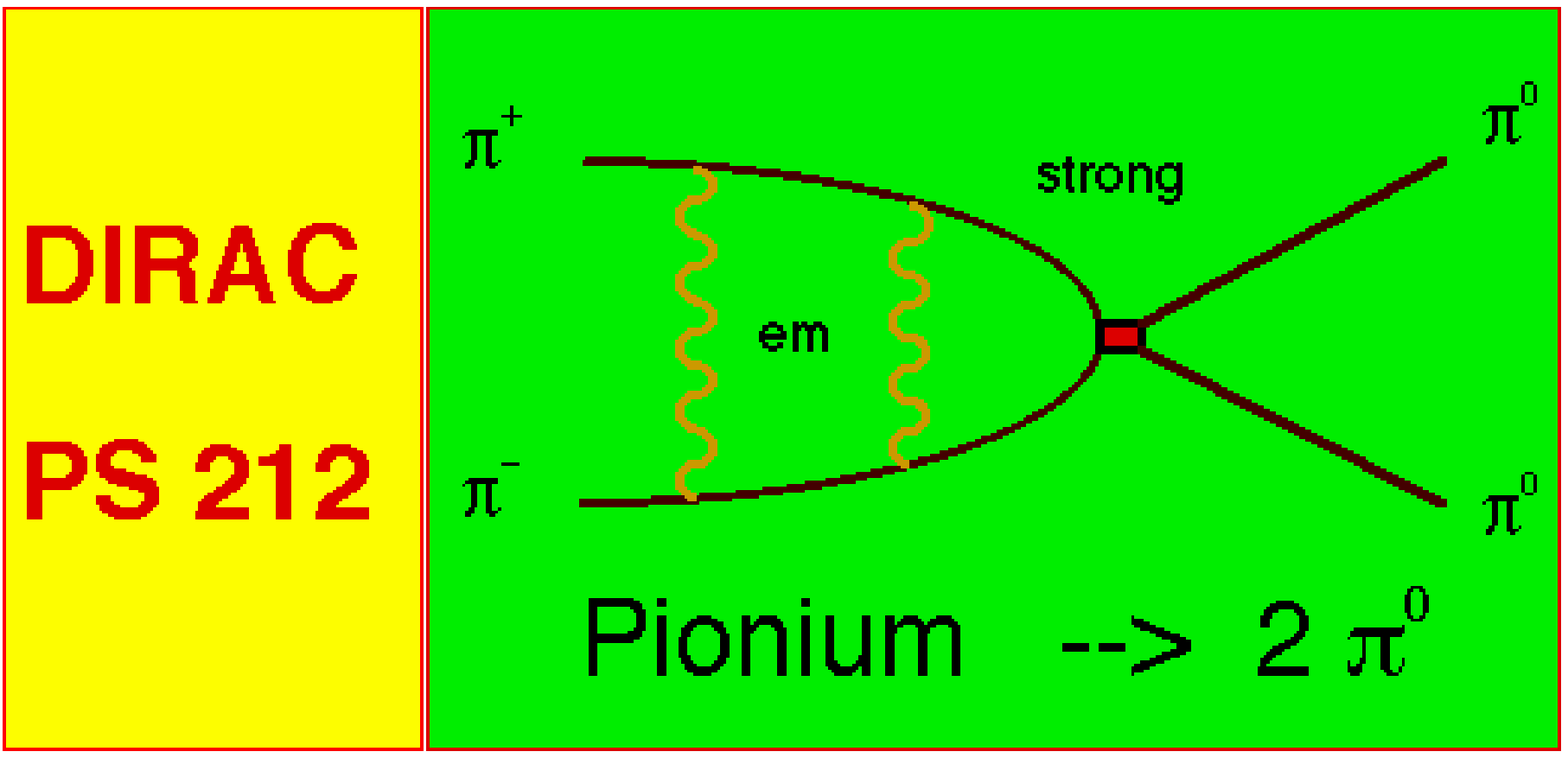,width=8.2cm,angle=00}\\
{\bf DIRAC Experiment at CERN:\\ 
LIFETIME MEASUREMENT OF PIONIUM\\}
\end{center}
\end{figure}

\begin{center}
Collaboration DIRAC\\
J. SCHACHER\\
\vspace*{0.2cm}
{\it Laboratorium fuer Hochenergiephysik, Universitaet Bern,} 
{\it CH-3012 Bern, Switzerland}
\end{center}


\vspace*{.5cm}
\begin{abstract}
{\small
The DIRAC experiment, a magnetic double arm spectrometer, 
aims to measure the $\pi^+\pi^-$ atom lifetime with $10\%$ 
precision, using the high intensity 24~GeV/c 
proton beam of the CERN Proton Synchrotron. Since the value 
of this lifetime of order $10^{-15}$~s is dictated by 
strong interaction at low energy, a precise measurement of 
this quantity enables to study characteristic pion parameters 
in detail and to submit predictions of QCD to a severe check. 
}
\end{abstract}

\vspace*{1cm}

INTRODUCTION. Pion scattering at low energies involves 
only the lightest observed hadrons and, 
therefore, is a key issue of low energy QCD. This 
pure process has been analysed 
in terms of scattering amplitudes and, hence, 
in terms of scattering lengths. By means of 
CHiral Perturbation Theory (CHPT) precise values for pion 
scattering lengths presumably at the $5\%$ level 
are and will be predicted~\cite{BIJ}, 
experimentally not confirmed up to now (the best precision: 
$\sim 20\%$ by Rosselet et al.~\cite{ROS}). 

The Collaboration DIRAC (DImeson Relativistic 
Atom Complex, PS212 at CERN) intends to measure the 
ground-state lifetime of the atom ``Pionium''. 
Pionium or $A_{2\pi}$ is the Coulomb bound state, 
formed by $\pi^+$ and $\pi^-$ mesons, with the following 
predicted properties:

\noindent
1. Bohr radius: $r_B=387$~fm
 
\noindent
2. Coulomb energy of the ground state: $E(1$S$)=-1.86$~keV
 
\noindent
3. Ground-state quantum numbers: $J^{PC}=0^{++}$ 

\noindent
4. Pionium will decay in more than $99\%$ of the cases by 
strong interaction into two neutral pions 
with a ground-state lifetime $\tau$ of about 3~fs. 

The $\pi^++\pi^-\rightarrow\pi^0+\pi^0$ reaction or 
decay rate from the atomic ground-state must be 
given by characteristic pion parameters at 
low energy around threshold, i.e. at 
$E=2m_\pi(1-\alpha^2/8)$. As a well defined quantity 
in pion physics, the pionium lifetime 
should be known with good precision. Therefore, 
DIRAC aims to extract from data 
at least a $10\%$ precise value for $\tau$. In the case of 
using Deser-type relationships~\cite{DES} 
between $\tau$ and $s$-wave scattering lengths 
(isoscalar minus isotensor) 
$\tau^{-1}=C\cdot \Delta^2$ 
with $\Delta=a_0-a_2$, 
we envisage a determination of 
scattering (or decay) lengths (difference) down 
to few percent~\cite{DIR}.

PRODUCTION, DETECTION AND LIFETIME MEASUREMENT OF 
$A_{2\pi}$. The method of pionium 
production, observation and 
lifetime measurement has been 
proposed many years ago by Nemenov, and details 
are described in reference~\cite{NEM}. 
Coulomb bound states or atoms can be produced 
in processes, where oppositely 
charged particles are emitted in the final state. 
Atoms, formed in this way, are in 
S-states. The corresponding cross section 
is proportional to the double inclusive 
production cross section $\sigma^0_s$ for $\pi^+\pi^-$ 
pairs from short-lived sources excluding Coulomb 
interaction in the final state and to the 
probability density function of pionium at the 
origin (with quantum numbers $n$ and $l=0$):

\begin{equation}\label{prod}
\frac{d\sigma^A_n}{d{\vec{p}}_A}  =
{(2\pi)}^3 \,
\frac{E_A}{M_A} \, |\Psi_n(0)|^2 \, \left.
\frac{d\sigma^0_s}{d{\vec{p}}_1 \, d{\vec{p}}_2}
\right|_{\vec{p}_1=\vec{p}_2=\vec{p}_A/2}  \,
\end{equation}

\noindent where $\vec{p}_A$, $E_A$ 
and $M_A$ are momentum, energy and mass 
of the $\pi^+\pi^-$ atom in the 
laboratory  system, respectively; 
${\vec{p}}_1$ and 
${\vec{p}}_2$ are the $\pi^+$ and 
$\pi^-$ momenta in the laboratory system.  
The $\pi^+$ and $\pi^-$ momenta obey the relation 
$\vec{p}_1=\vec{p}_2=\vec{p}_A/2$.

For the DIRAC experiment it is 
proposed to generate $A_{2\pi}$ atoms 
in 24~GeV/c proton~-~nucleus (e.g. Ti or Ni) 
collisions at the CERN PS. After production in a thin 
target of around $100~\mu$m thickness, 
the relativistic ($\gamma\simeq 15$) 
atoms may either decay into $\pi^0\pi^0$ or 
get excited or broken up (ionized) in the material of 
the that target. In the case of breakup, 
a characteristic charged pion pair 
will appear. The so-called ``atomic pair'' will 
be observed by recording pions, which show 
a low relative momentum in their 
centre of mass system  ($q< 3$~MeV/c). These pions have  
a small opening angle ($\theta_{+-}< 3$~mrad) 
and about the same laboratory energies 
($E_+\simeq E_-$ at the $0.3\%$ level). 
Pions from $A_{2\pi}$ breakup will lead to an excess of events 
above a large background of $\pi^+\pi^-$ pairs from 
free states (``free pairs'').

If the target thickness is of the order of the 
length for pionium interaction with target 
atoms, then the fraction of ionized pionia will 
be well measurable. The experimental 
setup, a magnetic double arm spectrometer, 
is designed to identify charged pions and to measure 
{\em relative pair momenta q} with high resolution of 
$\delta q\simeq 1$~MeV/c. By means of this precision 
apparatus the signal - the number of ``atomic pairs'' - can 
be separated from background - ``free pairs'' - and will 
be found in the following way: The number $n_A$ of signal events 
is given by the difference between the full number of 
$\pi^+\pi^-$ pairs in the low momentum 
range ($q< 3$~MeV/c), in which most of the  
``atomic pairs'' are expected to fall, and the computed 
number of ``free pairs'' in the same $q$ interval. To extract 
that number from the data, the distribution of the momenta $q$ 
for ``free pairs'' is fitted for $q> 3$~MeV/c 
by a function, based on the accidental
pion pair distribution, 
and then extrapolated back to the signal region. 
The fit function takes into account $\pi^+\pi^-$ pair 
production from short-lived (e.g. $\rho$) as well as 
from long-lived (e.g. $\eta_0$) sources (for 
details see~\cite{DIR}). The total number $N_A$ of generated 
$A_{2\pi}$ is obtained from the measured number of 
``free pairs'' originating from short-lived sources for 
$q< 3~$MeV/c and by using formula (1). Thus the 
probability for $A_{2\pi}$ breakup can be defined as 
$P_{br}(\gamma\cdot \tau)=n_A(\gamma\cdot \tau)/N_A(\gamma)$. 
For a given target material and thickness, this ratio depends 
on the {\bf pionium lifetime $\tau$}. The function 
$P_{br}(\gamma\cdot \tau)$ is given by 
the interaction of pionium with the target atoms
and is calculated with good precision. With help of 
this dependence the experimentally found breakup 
probability $P_{br}(exp)$ allows us to derive a value 
for the pionium lifetime.

\begin{figure}
\begin{center}
\epsfig{file=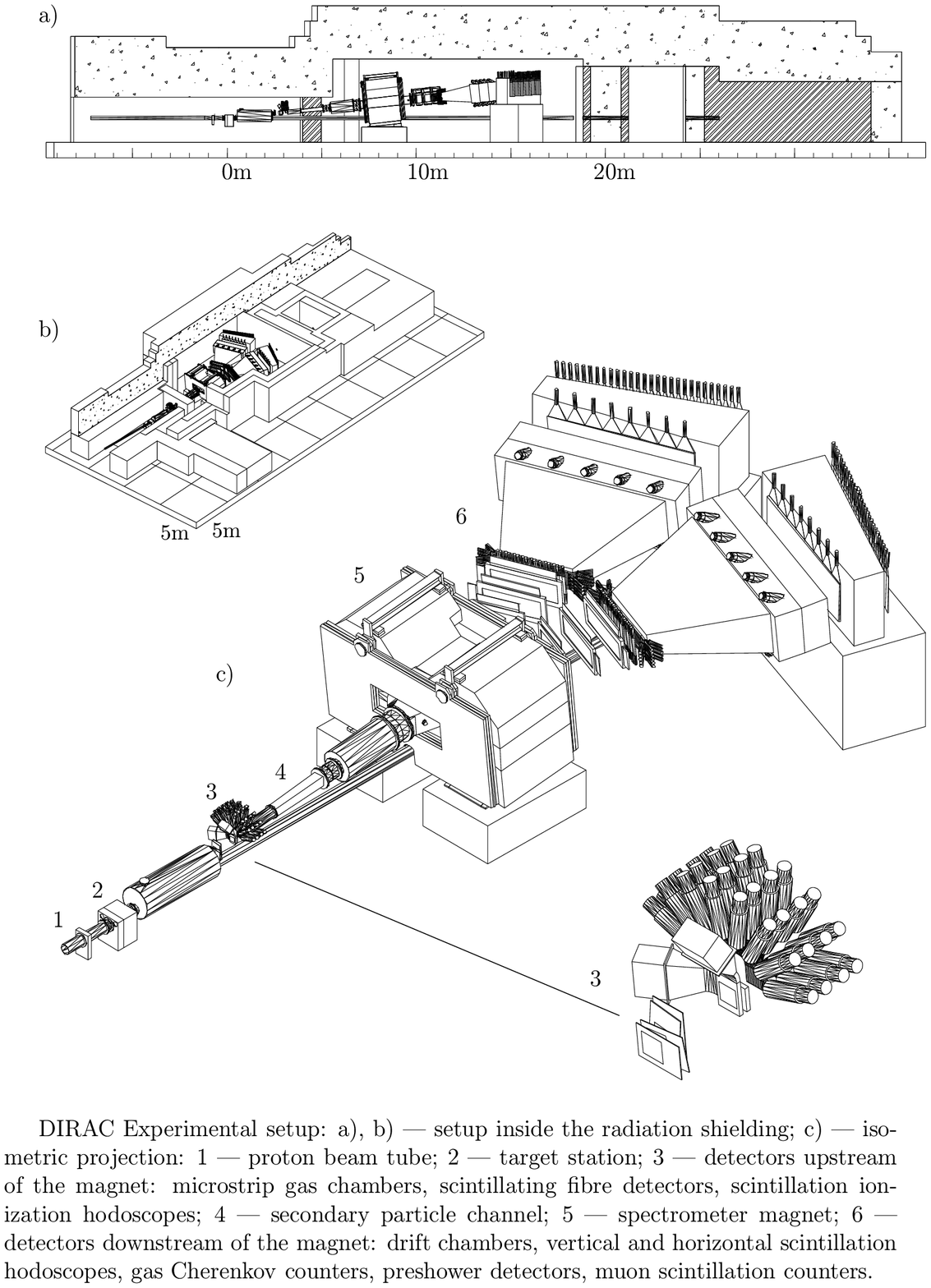,height=22cm}
\end{center}
\end{figure}

SETUP, TRIGGER AND PIONIUM YIELD. The DIRAC setup  
(see figure) is going to be installed in 
the ZT8 beam area of the PS East Hall just now. 
Extracted from PS the 24~GeV/c proton beam 
is focused on the target (2 in figure). The ZT8 beam 
will be operated at an intensity of about $2\cdot 10^{11}$ 
protons per spill of length 0.35~s with a cycle 
period of 14.4~s, corresponding to a $2.4\%$ duty cycle. 
The secondary particle channel with a aperture of 
1.2~msr is arranged at an angle of $5.7^0$ to the 
primary proton beam and will detect pairs of 
$\pi^+\pi^-$ in the pion momentum 
interval $1\div6$~GeV/c. 

DIRAC~\cite{DIR} consists of the following components: 
Four micro-strip gas chambers, a set 
of scintillating fibre detectors and  
scintillation hodoscopes (3 in figure) near the target, 
a spectrometer magnet (5) of 2.3~Tm bending power and 
two telescope arms (6), each equipped with drift 
chambers, vertical and horizontal 
scintillation hodoscopes, gas Cherenkov 
counters, preshower detectors and muon identifiers. The 
relative momentum resolution required for 
identification of ``atomic pairs'' is provided by the 
high coordinate resolution of the fibre detector, 
micro-strip gas chambers and the drift chambers. 
In order to reject electrons and muons, the system 
of threshold Cherenkov counters, operated with Nitrogen, 
preshower detectors and muon scintillation counters 
is being installed in each arm. For suppressing  
the large background rate a multilevel trigger logic 
is used. Besides a fast zero level trigger - a 2-arm 
coincidence $(VH\cdot Pr)_{1}\cdot (VH\cdot Pr)_{2}$ of 
vertical hodoscopes $VH$ and preshower detectors $Pr$ - 
the first level trigger is provided by 
a coincidence between the following responses of the two 
telescope arms: 
$$
(VH\cdot HH\cdot \overline{C}\cdot \overline{S_{\mu}})_{1}\cdot 
(VH\cdot HH\cdot \overline{C}\cdot \overline{S_{\mu}})_{2}
$$
where $VH$ and $HH$ mean the vertical and 
horizontal scintillation hodoscopes, $C$ the 
Cherenkov counters and $S_{\mu}$ the muon  
counters. 
At the higher levels the response 
of detectors upstream of the magnet 
is included. Special processors investigate 
the event topology on the basis of 
the pair opening angle (SFD) 
and the (double) ionization loss in the 
forward scintillation hodoscopes (FSD). 
The topological criterion is 
aimed to select events with 
low relative momentum of the 
detected particle pair. By these means the 
average rate of events recorded on 
tape is estimated to be around 30/s. 
To achieve the goal of DIRAC - a 
measurement of the $A_{2\pi}$ lifetime 
with $10\%$ precision - we have 
to consider a running time of at least 5 weeks 
(per target), corresponding to $\sim 20000$ 
recorded ``atomic pairs''. Of course additional 
data will be taken for tests, calibration and 
runs with different target materials.

CONCLUSION. From the experimental point of view, 
it is a challenge to produce in 
a high energy collision atomic states, 
in our case $\pi^+\pi^-$ atoms, and to measure 
their lifetime, which is of the order $10^{-15}$~s.

From the theoretical point of view, 
it is also a challenge to understand reliably 
the dependence of the $A_{2\pi}$ lifetime from 
characteristic pion parameters like scattering (decay) 
lengths and then to extract precise 
values for these quantities. 

\pagebreak

{\it Acknowledgments}.
I would like to thank the DIRAC Collaboration, especially 
L.L.~Nemenov, for many important discussions 
concerning our experiment. Furthermore I take the 
opportunity to acknowledge all the theoretical 
contributions to DIRAC by my Colleagues J. Gasser, 
H. Leutwyler and P. Minkowski from the Institute for 
Theoretical Physics at the University of Bern.
For the warm hospitality during the Workshop in Dubna I 
express my thanks to A.G. Rusetsky and his Colleagues. I am 
grateful to L.~Afanasyev for carefully reading my manuscript.

\end{document}